\documentclass{acm_proc_article-sp}

\usepackage{tabularx,array}
\usepackage{floatrow}
\usepackage[caption=false]{subfig}

\usepackage[Algorithm]{algorithm}
\usepackage[noend]{algpseudocode} 
\algrenewcommand\alglinenumber[1]{\tiny #1:}

\usepackage{amsthm}

\usepackage{hyperref}

\usepackage{comment}
\usepackage{enumitem}
\usepackage[utf8]{inputenc}

\usepackage{listings}

\newcolumntype{L}[1]{>{\raggedright\let\newline\\\arraybackslash\hspace{0pt}}m{
#1}}
\newcolumntype{C}[1]{>{\centering\let\newline\\\arraybackslash\hspace{0pt}}m{#1}
}
\newcolumntype{R}[1]{>{\raggedleft\let\newline\\\arraybackslash\hspace{0pt}}m{#1
}}

\lstdefinestyle{sparql}{
  basicstyle=\sffamily\scriptsize, 
  columns=fullflexible, 
  numbers=none,  
  frame=lines}

\lstdefinestyle{tiny}{
  basicstyle=\sffamily\tiny, 
  numbers=none,
  frame=none}
  
  \lstdefinestyle{query}{
  basicstyle=\sffamily\scriptsize, 
  numbers=none,
  frame=none}
  
\theoremstyle{definition}
\newtheorem{definition}{Definition}[section]

\newtheorem{example}{Example}[section]
\newtheorem{remark}{Remark}

 \hypersetup{
    colorlinks,%
    citecolor=black,%
    filecolor=black,%
    linkcolor=black,%
    urlcolor=black
}

\newcommand*\Let[2]{\State #1 $\gets$ #2}
\algrenewcommand\algorithmicrequire{\textbf{Precondition:}}
\algrenewcommand\algorithmicensure{\textbf{Postcondition:}}
\newcommand{\out}[1]{}
\begin{document}

\title{Modeling and Querying Data Cubes on the Semantic Web}

%
%
%
%
%

\numberofauthors{3} 
%
\author{
%
%
\alignauthor
Lorena Etcheverry\\
       \affaddr{Instituto de Computaci\'{o}n, Universidad de la Rep\'{u}blica}\\
       \affaddr{Montevideo, Uruguay}\\
       \email{lorenae@fing.edu.uy}
\alignauthor
Silvia Gomez\\
       \affaddr{Instituto Tecnol\'{o}gico de Buenos Aires}\\
       \affaddr{Buenos Aires, Argentina}\\
       \email{sgomez@itba.edu.ar}
\alignauthor
Alejandro Vaisman\\
       \affaddr{Instituto Tecnol\'{o}gico de Buenos Aires}\\
       \affaddr{Buenos Aires, Argentina}\\
       \email{avaisman@itba.edu.ar}
}

\pagenumbering{arabic}
\maketitle

\begin{abstract}
The web is changing the way in which data warehouses are designed, used, and 
queried. With the advent of initiatives such as Open Data and Open Government, 
organizations want to share their multidimensional data cubes and make them 
available to be queried online.  The RDF data cube vocabulary (QB),  the W3C 
standard to publish statistical data in RDF,  presents several limitations to 
fully support the multidimensional model. The QB4OLAP vocabulary extends QB to 
overcome these limitations, allowing  to implement the typical
OLAP operations, such as rollup, slice, dice, and  drill-across using 
standard SPARQL queries. In this paper we introduce a formal data model where the main object is the data cube, and 
define OLAP operations using this model, independent of the underlying 
representation of the cube.  We  show then that a cube expressed using  our model can be 
represented using the QB4OLAP vocabulary, and finally we provide a SPARQL implementation 
of OLAP operations over data cubes in QB4OLAP.
\end{abstract}


\section{Introduction}
\label{sec:intro}

On-Line Analytical Processing (OLAP) is a well-established approach for data 
analysis to support decision making, that typically relates to Data Warehouse 
(DW) systems. It is based on the multidimensional (MD) model which views data in 
an $n$-dimensional space, usually called a  data cube. There is a  large  number 
of MD models in the literature~\cite{Gomez2012,Hurtado1999,Vassiliadis1998}, 
based on the data cube metaphor. Historically, DW and OLAP have been used as 
techniques for data analysis within an organization, using mostly commercial 
tools with proprietary formats. However, initiatives like Open 
Data\footnote{\url{http://okfn.org/opendata/}} and Open 
Government\footnote{\url{http://opengovdata.org/}} are pushing organizations to 
publish MD data using standards and non-proprietary formats. Although several 
open source platforms for business intelligence (BI) have emerged in the last 
decade,   an open format to publish and share cubes among organizations is still 
missing.  The \textit{Linked Data}~\cite{Heath2011} initiative promotes sharing 
and  reusing data on the web using  \textit{semantic web} (SW) standards and 
domain ontologies expressed in the Resource Description Framework(RDF) (the 
basic data representation layer for the SW)~\cite{Klyne2004}, or in languages 
built on top of RDF (e.g., RDF-Schema~\cite{Brickley2004}). 

The need for tools and techniques allowing to publish 
and sharing data cubes did not take long to arise. 
Statistical data sets are usually published using the \textit{RDF Data Cube 
Vocabulary}\cite{Cyganiak2014} (also denoted 
QB), the current W3C standard. However, as we discussed in~\cite{Etcheverry2012a,Etcheverry2012b}, the QB vocabulary 
does not support dimension hierarchies and aggregate functions needed for OLAP 
analysis. To address this challenge, we proposed a new vocabulary called QB4OLAP~\cite{Etcheverry2012b}, 
that allows reusing data already published in QB just by adding the needed MD schema semantics (e.g., the hierarchical structure of the dimensions) and the corresponding instances that populate the 
dimension levels. Once a data cube is published using QB4OLAP, users are able  
to operate over it, not only through queries written in 
SPARQL~\cite{prud2011sparql} (the standard query language for RDF), but also 
using a high-level declarative OLAP language built taking advantage of the QB4OLAP 
metadata. 

In this paper we extend our previous work, presenting a formal data model for data cubes, and we use it to 
provide the semantics of a set of high-level OLAP operators over data cubes. 
We then show that a data cube represented using this model can be represented on 
the semantic web using the QB4OLAP vocabulary. Finally, we provide a SPARQL implementation of the OLAP operators over QB4OLAP.  

To put the reader in context, we next present the basic concepts on OLAP and SW data models.

\subsection{RDF  and the Semantic Web}
\label{sec:linked} 
 
Resource Description Framework (RDF) is a data model for expressing assertions over resources
identified by an internationalized resource identifier (IRI).   
Assertions are expressed as triples of the form  $(subject, predicate, object)$. 
A set of RDF triples or \emph{RDF data set}, can be seen as a
directed graph where the \emph{subject} and \emph{object} are nodes, and
the \emph{predicates} are arcs. Data values in RDF are called \emph{literals}.  
\emph{Blank nodes}   are used to represent anonymous resources or resources
without an IRI, typically  with  a structural function,  e.g., to group a set of
statements.   \emph{Subjects} are always resources or blank nodes, \emph{predicates} are always resources, 
and \emph{object} could be resources, blank nodes or literals.
A set of reserved words defined in RDF Schema 
(called the rdfs-vocabulary)\cite{Brickley2004} 
is used to define classes, properties, and to represent hierarchical relationships 
between them. For example, the triple (\emph{s, \texttt{rdf:type}, c}) explicitly states that \emph{s} is an instance 
of  \emph{c}  instance. Many formats for RDF serialization exist.  In 
this paper we  use Turtle~\cite{Beckett2011}.

SPARQL 1.1~\cite{prud2011sparql}  is the current  W3C standard query 
language for RDF. Its query evaluation  mechanism is based on 
subgraph matching: RDF triples are interpreted as nodes and edges of directed 
graphs, and the query graph is matched  to the data graph, instantiating  the variables 
in the query. The selection criteria is expressed using a graph pattern in the
\texttt{WHERE} clause.
Relevant to OLAP queries,  SPARQL supports aggregate functions and  the  
\texttt{GROUP BY} clause.

\subsection{OLAP}
\label{sec.olap}

Data Warehouses (DW)  integrate data from multiple sources, also keeping their 
history  for analysis and decision support. DWs represent data according to  
{\em dimensions} and {\em facts}. The former reflect the perspectives from 
which data are viewed, and we may have several of them. The latter corresponds 
to (usually)  quantitative data (also known as \emph{measures})  associated 
with  different dimensions. Dimensions organize elements, or members, in  
hierarchies, where each element belongs to a category (or level) in  a hierarchy. 
These are the main components of the multidimensional model,  
which represents facts in an $n$-dimensional space, 
usually called a {\em data cube}, whose axes are the  dimensions, and whose cells 
contain the values for the measures.

Online Analytical Processing (OLAP) is the process of querying a data cube, where
facts can be aggregated and disaggregated via operations called \textit{roll-up} and  \textit{drill-down}, respectively, and filtered 
through \textit{slice} and \textit{dice}, among other operations. 

As an illustration, the facts related to the sales of a company may be associated with 
the dimensions \textsf{Time} and \textsf{Location}, representing the sales at 
certain locations in certain periods of time. Assuming that facts (sales) are recorded at granularities 
month and city in dimensions \textsf{Time} and \textsf{Location}, respectively, a point in this space 
could be (January 2014, Buenos Aires), and the measure in this cell indicates the  
amount of the sales in January 2014, at the Buenos Aires branch. A \textit{roll-up} operation over dimension \textsf{Time}
up to level year would produce the yearly amount of sales for each city.

\subsection{Problem Statement and Contributions}
\label{sec:problem}

Ciferri et al.~\cite{CCG+13} have shown that, opposite to the usual belief, 
most of the multidimensional data models in the  literature are at the logical 
level rather than at a conceptual level, and that the data cube is far from 
being the focus of these models. Therefore, the authors sketched (quite 
informally) a model and algebra where the data cube is 
a first-class citizen.  Along the same lines, G\'{o}mez el 
al.~\cite{gomez2012generic} showed that such a 
model can be used to seamlessly query many kinds of multidimensional data 
(e.g., discrete and continuous geographic data). We follow these lines of 
thought, and, as our \textit{first contribution}, formalize a data model where 
the main object is the data cube over which a  conceptual 
query language where  the operators  manipulate the data cube, can be defined. 
This way, the user just queries data cubes, independently of the underlying 
data representation. Moreover, in our data model, the semantics of these 
operators is clearly defined using the notion of a lattice of cuboids, which is 
later used for query processing and rewriting. As a \textit{second contribution} 
we  show that a data cube represented using our data model can be published on 
the semantic web using the QB4OLAP vocabulary. Finally, as a \textit{third contribution}, taking advantage of the structural metadata provided by the QB4OLAP representation, we  present algorithms that produce a SPARQL implementation of the main OLAP operators over QB4OLAP data cubes.
That is, an OLAP user would not need to have any knowledge of SPARQL at all, and still be able to query cubes over the semantic web.

The remainder of this paper is organized as follows. 
Section \ref{sec:example} introduces our running example.
Section \ref{sec:cubes}  presents a formal data model for data cubes. 
Then, Section \ref{sec:QB4OLAPdc} describes the representation of data cubes in QB4OLAP.
Section~\ref{sec:queries} introduces a high-level OLAP query language purely based on operations over a data cube, and
presents a set of algorithms to produce a SPARQL implementation of these operations over QB4OLAP data cubes.
Finally, Section~\ref{sec:related} discusses related work, and Section~\ref{sec:conclusion} concludes the paper.

\section{Running Example}
\label{sec:example}

Throughout this paper we will be using an example based on statistical data 
about 
asylum  applications to countries in the European Union, provided by  
Eurostat\footnote{\url{
http://epp.eurostat.ec.europa.eu/cache/ITY_SDDS/EN/migr_asyapp_esms.htm}}. 
This dataset contains information about the number of asylum applicants by 
month, age, sex, citizenship, and country that receives the application. It
is published using QB in the Eurostat - Linked Data 
dataspace\footnote{\url{http://eurostat.linked-statistics.org/}}. Basically, 
a QB dataset  is composed of a set of \textit{observations} that represent 
data instances that adhere to a schema, represented by a \textit{data structure 
definition} (DSD). However, QB lacks of the capability to represent
dimension aggregation hierarchies. To allow OLAP operations,  QB4OLAP 
allows building cube 
schemas on top of  the \textit{observations} already published using QB. In 
this way,  the 
cost of adding OLAP capabilities to existing datasets is the cost of 
building the new dimension schema (the analysis 
dimensions), and populating its instances.  In this   example 
we built simple 
dimension hierarchies to organize   countries  into continents, 
and months into years. 

\begin{figure}
\centering
\includegraphics[width=\linewidth]{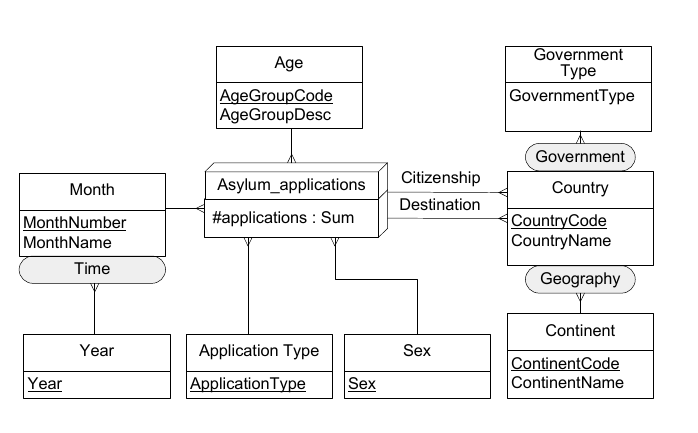}
\caption {Conceptual schema of the Asylum Applications cube}
\label{fig.conceptual}
\end{figure}

Figure~\ref{fig.conceptual} shows the conceptual 
schema of the  data cube (extended with hierarchies), 
using the MultiDim notation \cite{VZ14}. 
The \textsf{asylum\_applications} cube has  a measure
 (\textsf{\#applications}) that represents the number of applications. 
This measure can be analyzed according to six  dimensions: the 
\textsf{sex}  of the applicant, \textsf{age} which organizes applicants 
according to their 
age group, \textsf{time} which represents the time of the application and 
consists of two 
levels (month and year), \textsf{application\_type} that represents if the 
applicant is a first-time applicant 
or a returning  applicant, and a geographical dimension that organizes 
countries 
into continents (\textsf{Geography} hierarchy) or according to its government 
type (\textsf{Government} hierarchy). 
This geographical dimension participates in the cube with two different roles: 
as 
the \textsf{citizenship} of the asylum applicant, 
and as  the \textsf{destination} country of the application.

\section{Data Cubes}
\label{sec:cubes}

In multidimensional models, data  are organized as   cubes 
whose axes are \textit{dimensions}.
Each point in this multidimensional space is mapped  into
one or more spaces of \textit{measures}.  
Dimensions are organized in \textit{hierarchies}  that allow
analysis at different aggregation \textit{levels}. The values in a dimension level 
are called \textit{members}, and may have properties or \textit{attributes}. 
 Members  in a    
level must have a corresponding  member in the upper level in the hierarchy, and
this correspondence is defined through so-called rollup functions. 
 In this section we present a formal model for data cubes upon which we build 
 our query language. 

\subsection{Data Cubes Formalization}
\label{sec:cubes:formal}
\vspace{0.2cm}
\begin{definition}(\textbf{Dimension schema}). A \textit{dimension\\
schema} is a tuple 
$\langle \mathsf{nD}, \mathcal{L}, \rightarrow, \mathcal{H}\rangle$ where:
(a) \textsf{nD} is the name of the dimension; 
(b) $\mathcal{L}$ is a set of tuples $\langle \mathsf{name}_l, \mathsf{A}_l 
\rangle$, called \textit{levels}, where $\mathsf{name}_l$ identifies 
a level in $\mathcal{L}$, and $\mathsf{A}_l= \langle a_1, \dots, a_n \rangle$ 
is a tuple
of level \textit{attributes}. Each attribute $a_i$ has a domain $Dom(a_i)$ with 
$1 \leq i \leq n$; 
(c) $(\mathcal{L}, \rightarrow)$ represents a lattice with a unique bottom 
level and a unique top level (\textsf{All}), where `$\rightarrow$'  
is a partial order that defines a  
\textit{parent-child} relation between pairs of levels in $\mathcal{L}$; 
(d) $\mathcal{H}$ is a set of tuples $\langle \mathsf{name}_h, \mathsf{L}_h 
\rangle$, called \textit{hierarchies}, where $\mathsf{name}_h$ identifies the 
hierarchy, 
and $\mathsf{L}_h \subseteq \mathcal{L}$ is the set of levels that participate 
in the hierarchy.\qed
\label{def:dimSc}
\end{definition}

\begin{remark} All  levels in a dimension schema must belong to at least one 
hierarchy. For each dimension $\mathsf{nD}$ with  schema $\langle \mathsf{nD}, 
\mathcal{L}, \rightarrow, \mathcal{H}\rangle$, where
$\mathcal{H} = \{ \langle \mathsf{h}_1, \mathsf{L}_{h_1} \rangle, \dots, 
\langle 
\mathsf{h}_i, \mathsf{L}_{h_i} \rangle \}$, then
$\bigcup_{j=0}^{j=i}\mathsf{L}_{h_j} = \mathcal{L}$.

In addition, all the dimensions contain at least one default hierarchy $\langle 
\mathsf{name}_h, \mathsf{L}_h 
\rangle$, with $\mathsf{L}_h = \mathcal{L}$.
\qed
\end{remark}
 
\begin{example} (\textsf{Citizenship} dimension schema) The schema of the 
\textsf{Citizenship} dimension   in Figure \ref{fig.conceptual} is 
defined as: \\
$\langle \mathsf{Citizenship}$
$\mathcal{L}= \{ \langle \mathsf{Country},\langle 
\mathsf{countryCode},\mathsf{countryName}\rangle \rangle,$\\
$\langle \mathsf{Continent},\langle \mathsf{continentCode},\mathsf{continentName}\rangle \rangle, $\\
$\langle \mathsf{GovernmentType},\langle \mathsf{governmentType}\rangle \rangle, $\\
$\langle \mathsf{All},\langle \mathsf{all}\rangle \rangle \}$; \\
$ \mbox{`}\rightarrow\mbox{'} = \{ \mathsf{Country} \rightarrow \mathsf{Continent},\mathsf{Country} \rightarrow \mathsf{GovernmentType},$\\
$\mathsf{Continent} \rightarrow \mathsf{All},\mathsf{GovernmentType} \rightarrow \mathsf{All} \};$\\
$\mathcal{H} = \{ \langle \mathsf{Geography}, \{ \mathsf{Country},\mathsf{Continent},\mathsf{All} \} \rangle, $\\
$\langle \mathsf{Government}, \{\mathsf{Country},\mathsf{GovernmentType},\mathsf{All} \} \rangle \} $ \\
\qed
\label{ex:dimSc} 
\end{example}

\begin{definition} (\textbf{Dimension instance}). A \textit{dimension instance} 
 for  a dimension schema $\langle \mathsf{nD}, \mathcal{L}, \rightarrow,  
\mathcal{H} \rangle$ 
is a tuple
$\langle \langle \mathsf{nD}, \mathcal{L}, \rightarrow,  \mathcal{H} \rangle, 
\mathcal{T}_L, 
\mathcal{R} \rangle$ where:
 (a)   $\mathcal{T}_L$ is a  finite set of tuples of the form $\langle v_1, v_2, 
\dots, v_{n_l} \rangle~\forall \mathsf{L}=\langle \mathsf{L}, \langle a_1, 
\dots, a_n \rangle  \rangle \in \mathcal{L} $
 such that $\forall i, i=1,\dots,n, v_i \in Dom(a_i)$; 
(b)   $\mathcal{R}$ is a finite set of relations 
$\mathsf{RUP}^{\mathsf{L}_j}_{\mathsf{L}_i}, \mathsf{L}_i, \mathsf{L}_j \in \mathcal{L}$, 
 such that $\mathsf{L}_i \rightarrow \mathsf{L}_j~\in~\mathcal{L}, \rightarrow)$. Each relation 
$\mathsf{RUP}^{\mathsf{L}_j}_{\mathsf{L}_i}$ relates members of $\mathsf{L}_i$ 
(child level) with members of $\mathsf{L}_j$ (parent level).
 \qed
\label{def:dimIns}
\end{definition}

 \out{
 \begin{remark}
 Most multidimensional models assume that relations between parent and child 
 levels are functions. This restriction only allows to represent some cases, 
 where each  member 
 in the child level has one associated member  in the parent level. By using 
 relations we relax this restriction,  thus supporting non strict 
 hierarchies~\cite{VZ14}.
 \end{remark}
}

\begin{example} (\textsf{Citizenship} dimension instance) A possible instance 
of  the \textsf{Citizenship} dimension  in Figure \ref{fig.conceptual} 
is:\\ 
$ \mathcal{T}_{\mathsf{Country}}= \{ \langle $ \textsf{'AD', 
'Andorra'} $\rangle, 
\dots,\langle $\textsf{'ZW', 'Zimbabwe'}$\rangle \} $ \\
$ \mathcal{T}_{\mathsf{Continent}} = \{ \langle $\textsf{'AF', 
'Africa'}$\rangle, \dots,\langle$\textsf{'OC', 'Oceania'}$\rangle \} $\\
$ \mathcal{T}_{\mathsf{GovernmentType}} = \{ \langle 
$\textsf{'Republic'}$\rangle, \dots,\langle$\textsf{'Unitary state'}$\rangle \} 
$\\
$ \mathcal{T}_{\mathsf{All}}= \{ \langle$ \textsf{'all'}$\rangle \} $\\
$ \mathcal{R}= \{ \mathsf{RUP}_{\mathsf{Country}}^{\mathsf{Continent}}, 
\mathsf{RUP}_{\mathsf{Continent}}^{\mathsf{All}},$
$ \mathsf{RUP}_{\mathsf{Country}}^{\mathsf{GovernmentType}}, \\
\mathsf{RUP}_{\mathsf{GovernmentType}}^{\mathsf{All}} \}$, with 
$ \mathsf{RUP}_{\mathsf{Country}}^{\mathsf{Continent}} = \{ (\langle $ 
\textsf{'AD', 'Andorra'}$\rangle$, $\langle$\textsf{'EU', 'Europe'}$\rangle ), 
\ldots,$
$ (\langle $ \textsf{'ZW', 'Zimbabwe'}$\rangle$, $\langle$\textsf{'AF', 
'Africa'}$\rangle ) \};$ 
$ \mathsf{RUP}_{\mathsf{Country}}^{\mathsf{GovernmentType}} = \{ (\langle $ 
\textsf{'AD', 'Andorra'}$\rangle$, $\langle$\textsf{'Unitary state'}$\rangle ), 
\ldots,$  
$ (\langle $ \textsf{'ZW', 'Zimbabwe'}$\rangle$, $\langle$\textsf{'Presidential 
system'}$\rangle ) \}$;\\
$ \mathsf{RUP}_{\mathsf{Continent}}^{\mathsf{All}} = \{ (\langle x \rangle$, 
$\langle$\textsf{all}$\rangle)\ | x \in \mathcal{T}_{\mathsf{All}}  \}$;\\ 
$ \mathsf{RUP}_{\mathsf{GovernmentType}}^{\mathsf{All}} = \{ (\langle x 
\rangle$, $\langle$\textsf{all}$\rangle)\ | x \in \mathcal{T}_{\mathsf{All}}  
\}.$\\ 
\qed
\label{ex:dimIns}
\end{example}

\begin{definition} (\textbf{Cube schema}). A \textit{cube schema} is a tuple 
$\langle \mathsf{nC}, \mathcal{D}, \mathcal{M} , \mathcal{F} \rangle$ where:
  (a)  $\mathsf{nC}$ is the name of the cube; 
  (b) $\mathcal{D}$ is a finite set of dimension schemas (see Def. 
\ref{def:dimSc}); 
 (c)  $\mathcal{M}$ is a finite set of attributes, where each   $m \in 
 \mathcal{M}$, called \textit{measure}, has domain $Dom(m)$; 
 (d) $\mathcal{F}:\mathcal{M} \rightarrow \mathcal{A}$ is a function that 
maps each measure in $\mathcal{M}$ to an aggregate function in $\mathcal{A}$. 
 \qed
\label{def:cubeSch}
\end{definition}

\begin{example} (\textsf{Asylum\_application} cube schema) We define the cube 
schema, presented in Figure \ref{fig.conceptual} as:  \\ 
$ \langle$ \textsf{Asylum\_application}, 
\{\textsf{Sex},\textsf{Age},\textsf{Time},\textsf{Application\_type},\\
 \textsf{Citizenship}, \textsf{Destination}\}, \{\textsf{\#applications}\}, \\ 
 \{\textsf{\#applications,Sum}\}  $\rangle$, where dimension \textsf{Citizenship} is defined as in Example 
\ref{ex:dimSc}. 
We  omit the definition of the other dimensions, for the sake of brevity and to avoid redundancy. 
\qed
\label{ex:cubeSch}
\end{example}

To define a cube instance we need to introduce the notion of cuboid.

\begin{definition} (\textbf{Cuboid instance}). Given: (a) a cube schema 
$\langle \mathsf{nC}, \mathcal{D}, \mathcal{M}, \mathcal{F} \rangle$, where $|\mathcal{D}|= 
D$ and $|\mathcal{M}|= M$, (b) a dimension instance $\mathsf{I}_i$ 
for each $\mathsf{D}_i \in \mathcal{D}, \forall i, i=1,\dots,D$, and (c) a set 
of levels $\mathcal{V}_{Cb}=\lbrace l_1, l_2, \dots, l_D \rbrace$ 
where $l_i \in \mathcal{L}_i$ of $\mathsf{D}_i \in \mathcal{D}, \forall i, 
i=1,\dots,D$, such that there are not two levels belonging to the same 
dimension, a \textit{cuboid instance} \textsf{Cb} is a partial function
$\mathsf{Cb}:\mathcal{T}_{l_1} \times \dots \times \mathcal{T}_{l_D} 
\rightarrow Dom(m_1)\times \dots \times Dom(m_M)$, where $m_k \in \mathcal{M}, 
\forall k, k=1,\dots, M$. 
The elements in the domain of \textsf{Cb} are called \textit{cells}, and 
$\mathcal{V}_{Cb}$ it the \textit{set of levels} of the cuboid.
\qed
\label{def:cuboidInstance}
\end{definition}

\begin{example} (Cuboid instance) Consider the cube schema 
\textsf{Asylum\_application} defined in Example \ref{ex:cubeSch}. A possible
instance of the cuboid $\mathsf{Cb}_1$, where
$\mathcal{V}_{\mathsf{Cb}_1}=$\{\textsf{Sex}, \textsf{Age}, \textsf{Month}, \textsf{Application\_type},
 \textsf{Country}, \textsf{Country}$\} $ is presented in Figure 
\ref{fig:cuboidInstance}, using a tabular representation, where the first row 
lists the dimensions in the cuboid, and the second row lists the dimension 
level corresponding to the cuboid. 
\vspace{-0.5cm}\qed
\label{ex:cuboidInstance}
\end{example}

\begin{figure*}
\caption{Tabular representation of a cuboid instance of the 
\textsf{Asylum\_application} cube schema}
\tiny
\begin{tabular}{|c|c|c|C{2cm}|C{4.5cm}|C{2.5cm}|C{1.5cm}|}
\hline
\textbf{Sex} & \textbf{Age} & \textbf{Time} &\textbf{Application\_type} & 
\textbf{Citizenship} & \textbf{Destination} & \textbf{Measures} \\ \hline
\textit{Sex} & \textit{Age} & \textit{Month} & \textit{Application\_type} & 
\textit{Country} & \textit{Country} &\textit{ \#applications} \\ \hline
M & 14 to 17 & 201301, January 2013 & new applicant & CM, Cameroon & BE, Belgium & 5 \\ \hline
F & less than 14 & 201303, March 2013 & new applicant & CM, Cameroon & FR, France & 5 \\ \hline
M & 18 to 34 & 201301, January 2013 & new applicant & CM, Cameroon & FR, France & 10 \\ \hline
F & 18 to 34 & 201301, January 2013 & new applicant & CD, Democratic Republic of the Congo & BE, Belgium & 25 \\ \hline
F & 18 to 34 & 201303, March 2013 & new applicant & CD, Democratic Republic of the Congo & BE, Belgium & 30 \\ \hline
\end{tabular}
\label{fig:cuboidInstance}
\end{figure*}

\begin{figure*}
\caption{Two cuboid instances of the \textsf{Asylum\_application} cube 
schema}
\centering
\tiny
\begin{tabular}{c}
\subfloat[Cuboid $\mathsf{Cb}_2$]{
\begin{tabular}{|c|c|c|C{2cm}|C{2cm}|C{2.5cm}|C{1.5cm}|}
\hline
\textbf{Sex} & \textbf{Age} & \textbf{Time} &\textbf{Application\_type} & 
\textbf{Citizenship} & \textbf{Destination} & \textbf{Measures} \\ \hline
\textit{Sex} & \textit{Age} & \textit{Year} & \textit{Application\_type} & 
\textit{Continent} & \textit{Country} &\textit{ \#applications} \\ \hline
M & 14 to 17 & 2013 & new applicant & AF, Africa & BE, Belgium & 5 \\ \hline
F & less than 14 & 2013 & new applicant & AF, Africa & FR, France & 5 \\ \hline
M & 18 to 34 & 2013 & new applicant & AF, Africa & FR, France & 10 \\ \hline
F & 18 to 34 & 2013 & new applicant & AF, Africa & BE, Belgium & 55 \\ \hline
\end{tabular}
\label{fig:cubInst1}
}\\
\subfloat[Cuboid $\mathsf{Cb}_3$]{
\begin{tabular}{|c|c|c|C{2cm}|C{2cm}|C{2.5cm}|C{1.5cm}|}
\hline
\textbf{Sex} & \textbf{Age} & \textbf{Time} &\textbf{Application\_type} & 
\textbf{Citizenship} & \textbf{Destination} & \textbf{Measures} \\ \hline
\textit{All} & \textit{Age} & \textit{Year} & \textit{Application\_type} & 
\textit{Continent} & \textit{Country} &\textit{ \#applications} \\ \hline
all & 14 to 17 & 2013 & new applicant & AF, Africa & BE, Belgium & 5 \\ \hline
all & less than 14 & 2013 & new applicant & AF, Africa & FR, France & 5 \\ \hline
all & 18 to 34 & 2013 & new applicant & AF, Africa & FR, France & 10 \\ \hline
all & 18 to 34 & 2013 & new applicant & AF, Africa & BE, Belgium & 55 \\ \hline
\end{tabular}
\label{fig:cubInst2}
}
\end{tabular}
\label{fig:Notation}
\end{figure*}

The sets of the cuboid instances that refer to the same cube schema, can be 
organized using the concepts of \textit{adjacent cuboids} and \textit{order 
between cuboids},   defined as follows.

\begin{definition} (\textbf{Adjacent Cuboids}). Two cuboids $\mathsf{Cb}_1$ and 
$\mathsf{Cb}_2$, that refer to the same cube schema, 
are \textit{adjacent} if their corresponding level sets $\mathcal{V}_{Cb_1}$ 
and $\mathcal{V}_{Cb_2}$ differ in exactly one level, i.e., $|\mathcal{V}_{Cb_1} 
-\mathcal{V}_{Cb_2}| = |\mathcal{V}_{Cb_2} -\mathcal{V}_{Cb_1}| = 1$. 
\qed 
\label{def:cuboidAdj}
\end{definition}

\begin{example} (Adjacent cuboids) Consider the cube schema defined in Example 
\ref{ex:cubeSch} and the cuboids
$\mathsf{Cb}_1,\mathsf{Cb}_2$, and $\mathsf{Cb}_3$ given by 
$\mathcal{V}_{\mathsf{Cb}_1}=$\{\textsf{Sex}, \textsf{Age}, \textsf{Month},  
\textsf{Application\_type},
\textsf{Country}, \textsf{Country}$\}$, 
$\mathcal{V}_{\mathsf{Cb}_2}=$\{\textsf{Sex}, \textsf{Age}, \textsf{Year}, \textsf{Application\_type},
\textsf{Country},\textsf{Country}$\} $ and
 $\mathcal{V}_{\mathsf{Cb}_3}=$\{\textsf{Sex}, \textsf{Age}, \textsf{Year},  \textsf{Application\_type},
\textsf{Country}, \textsf{Continent}$\} $. According to Definition 
\ref{def:cuboidAdj}, $\mathsf{Cb}_1$ is adjacent to $\mathsf{Cb}_2$, and 
$\mathsf{Cb}_2$ is adjacent to $\mathsf{Cb}_3$, but
$\mathsf{Cb}_1$ is not adjacent to $\mathsf{Cb}_3$.
\qed
\label{ex:cuboidAdj}
\end{example}

\begin{definition} (\textbf{Order between Adjacent Cuboids}). \newline
Given two adjacent cuboids 
$\mathsf{Cb}_1$ and $\mathsf{Cb}_2$, such that $\mathcal{V}_{Cb_1} 
-\mathcal{V}_{Cb_2} = \lbrace l_c \rbrace$ and $\mathcal{V}_{Cb_2} 
-\mathcal{V}_{Cb_1} = \lbrace l_p \rbrace$, and $l_p$ and $l_c$ are levels of 
the lattice $(\mathcal{L}, \rightarrow)$ 
of the dimension $D_k$ such that $l_c \rightarrow l_p$, then $\mathsf{Cb}_1 
\preceq \mathsf{Cb}_2$. \newline
Moreover, for each pair of adjacent cuboids $\mathsf{Cb}_1 \preceq 
\mathsf{Cb}_2$ each cell $c = (c_1, \dots, c_{k-1},c_k, c_{k+1}, \dots, c_n, 
m_1,m_2, \dots m_s) \in \mathsf{Cb}_2$ 
can be obtained from the cells in $\mathsf{Cb}_1$ as follows. \newline
Let $(c_1, \dots, c_{k-1},b_{k1}, c_{k+1}, \dots, c_n, m_{1,1},m_{2,1}, \dots 
m_{s,1})$ , \newline $(c_1, \dots, c_{k-1},b_{k2}, c_{k+1}, \dots, c_n, 
m_{1,2},m_{2,2}, \dots m_{s,2})$, \newline
$(c_1, \dots, c_{k-1},b_{kp}, c_{k+1}, \dots, c_n, m_{1,p},m_{2,p}, \dots 
m_{s,p})$  be cells in $\mathsf{Cb}_1$ where $(b_{k_i},c_k) \in RUP_{l_{k_c}}^{l{k_p}}, i=1\dots q$, that means that all  members $b_{k_i}$ in level $l_{k_c}$ 
in dimension $D_k$ are in a parent-child relation with the element $c_k$ in level 
$l_{k_p}$ in that dimension, and the measures in cell $c \in \mathsf{Cb}_2$ are 
obtained as $m_i = AGG_i(m_{i,1},\dots,m_{i,j})$, where $AGG_i$ is the 
aggregation function related to measure $m_i$.
\qed
\label{def:cuboidOrder}
\end{definition}

\begin{example} (Order between cuboids) Consider the cuboids 
$\mathsf{Cb}_1,\mathsf{Cb}_2$, and $\mathsf{Cb}_3$ in Example 
\ref{ex:cuboidAdj}. Then $\mathsf{Cb}_1 \preceq \mathsf{Cb}_2$, because 
\textsf{Month} $\rightarrow$ \textsf{Year} holds, and $\mathsf{Cb}_2 \preceq 
\mathsf{Cb}_3$, because \textsf{Country} $\rightarrow$ \textsf{Continent} holds.
\qed
\label{ex:cuboidOrder}
\end{example}

Finally we define a \textit{cube instance} as the lattice of all  possible 
cuboids that share the same cube schema.

\begin{definition} (\textbf{Cube Instance}). Given a cube schema \newline
$\langle \mathsf{nC}, \mathcal{D}, \mathcal{M}, \mathcal{F} \rangle$, 
where $|\mathcal{D}|= 
D$ and $|\mathcal{M}|= M$, and a dimension instance $I_i$ for each $D_i \in 
\mathcal{D}, i=1,\dots,D$, a cube instance $\mathsf{CI}$ is the lattice 
$\lbrace \mathsf{CB},\preceq \rbrace$ where \textsf{CB} is the set of all  
possible cuboids, and $\preceq$ is the order between adjacent cuboids in 
\textsf{CB}.
\qed 
\label{def:cubeIns}
\end{definition}

\begin{example}(Cuboids of \textsf{Asylum\_application}) Consider the cube 
schema defined in Example \ref{ex:cubeSch}. All possible combinations of 
the  levels in the six dimensions of 
the cube, lead to 216 cuboids, which are organized in a lattice. Assuming the 
instance of cuboid $\mathsf{Cb}_1$ in Figure~\ref{fig:cuboidInstance},  
Figures 4a and  4b  present 
tabular representations of  instances of cuboids 
$\mathsf{Cb}_2$, and $\mathsf{Cb}_3$ given by 
$\mathcal{V}_{\mathsf{Cb}_2}=$\{\textsf{Sex}, \textsf{Age}, \textsf{Year},  
\textsf{Application\_type}, \textsf{Continent}, \textsf{Country}$\} $, and 
 $\mathcal{V}_{\mathsf{Cb}_3}=$\{\textsf{All}, \textsf{Age}, \textsf{Year},  
\textsf{Application\_type}, \textsf{Continent},\textsf{Country}$\} $.
\qed
\end{example}

\section{QB4OLAP and  Data Cubes}
\label{sec:QB4OLAPdc}

In this section we first present QB4OLAP distinctive features, and then we describe in detail how the formal model defined in Section
\ref{sec:cubes} can be represented in QB4OLAP.

\subsection{QB4OLAP}
\label{sec:QB4OLAPdc:QB4OLAP}

\begin{figure*}[t]
\centering
\includegraphics[width=0.7\textwidth]{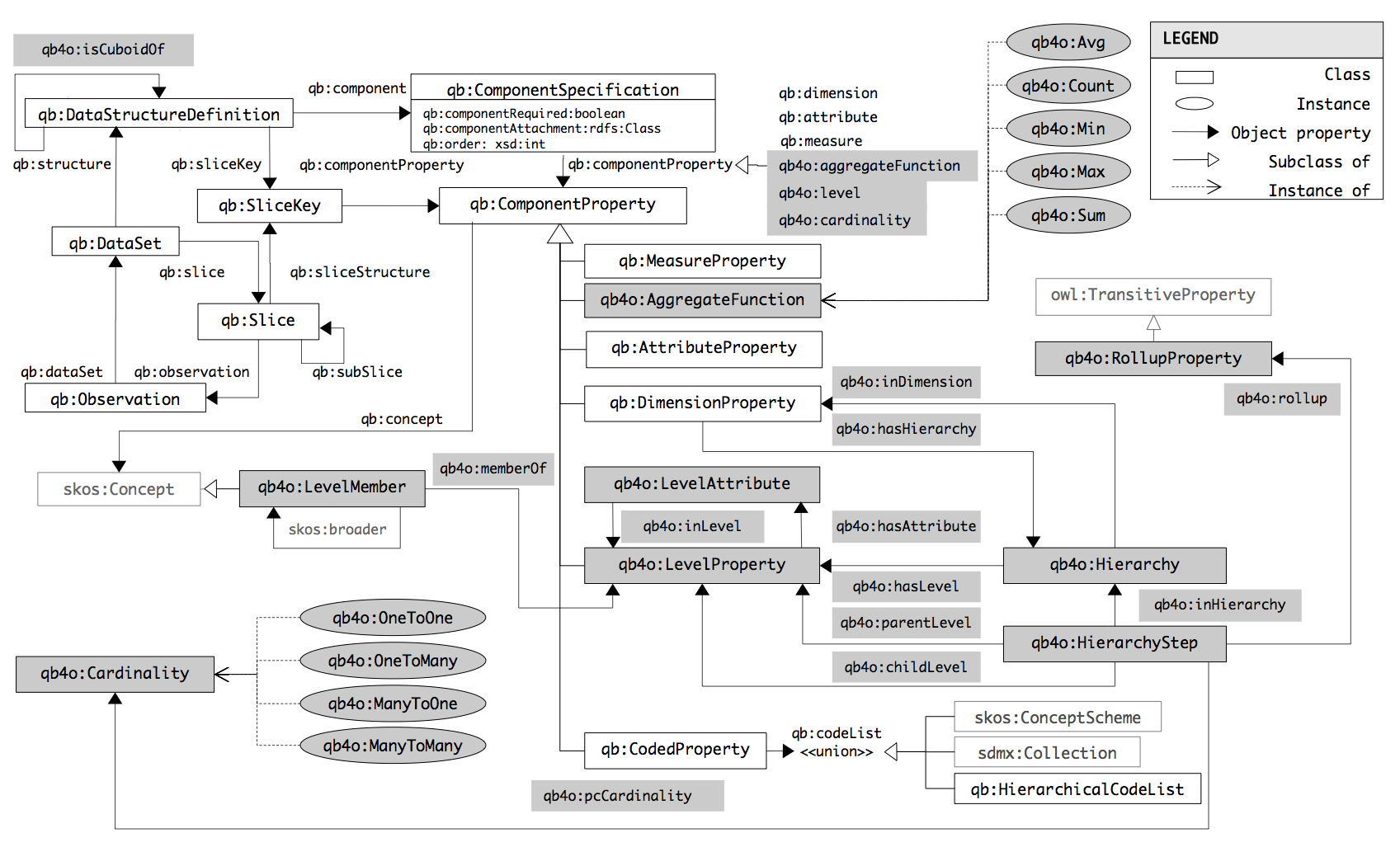}
\caption {QB4OLAP vocabulary (version 1.3)}
\label{fig:qb4olap}
\end{figure*}

QB4OLAP\footnote{\url{http://purl.org/qb4olap/cubes}} extends QB with a set of 
RDF terms that allow  representing the most common features of the  MD 
model. Figure~\ref{fig:qb4olap} depicts the QB4OLAP vocabulary. 
Original QB terms are prefixed with ``\texttt{qb:}'', while QB4OLAP terms are
prefixed with ``\texttt{qb4o:}'' and displayed in 
gray background. Capitalized terms represent RDF 
classes, non-capitalized terms represent RDF properties, and capitalized terms 
in italics represent class instances. An arrow from class $A$ to class $B$, 
labeled  $rel$ means that $rel$ is 
an RDF property with domain $A$ and range $B$. White triangles represent 
sub-class or 
sub-property relationships. Black diamonds represent \texttt{rdf:type} 
relationships (instances).
The range of a property can also be denoted using ``:''.
 
The rationale behind QB4OLAP includes: 
\begin{itemize}
\item QB4OLAP must be able to represent the most common features of the  MD 
model. The features considered are based on the MultiDim model~\cite{VZ14}. 
\item QB4OLAP must include all the metadata needed to implement 
OLAP operations as SPARQL queries. In this way,  OLAP users do not need to know 
SPARQL (which is the case of typical OLAP users), and even  wrappers for  OLAP 
tools could be developed to query  RDF data sets directly. 
We comment on this issue at the end of this section.
\item   QB4OLAP must allow to operate over already published observations  which 
conform to DSDs defined in QB, without the need 
of rewriting the existing observations, and with the minimum possible effort. 
Note that in a typical MD model, dimensions are usually orders of magnitude 
smaller than observations, which are the largest part of the data.
\end{itemize}

\subsection{Implementing data cubes in QB4OLAP }
\label{sec:QB4OLAPdc:cubes}

We next  sketch  how each of the concepts introduced  in Section 
\ref{sec:cubes:formal} can be represented in QB4OLAP.  The formal 
proof is outside the scope of this paper. We assume the reader
has basic knowledge of 
RDF syntax. 

\begin{definition}(\textbf{Dimension schema in QB4OLAP})
 A dimension schema in QB4OLAP is an RDF graph  that 
 uses terms defined in the QB and QB4OLAP vocabularies as follows:
 \vspace{-0.2cm} 
 \begin{itemize}
 \item Dimensions are defined as instances of the class \\
 \texttt{qb:DimensionProperty}.

 \item  Levels are defined as properties, which are instances of the class 
  \texttt{qb4o:LevelProperty}.

 \item Level attributes are defined as properties, which are instances of the class
  \texttt{qb4o:LevelAttribute}, and related to levels with the property \texttt{qb4o:inLevel} or its inverse   \texttt{qb4o:hasAttribute}.

  \item Hierarchies are defined as instances of the class \\
 \texttt{qb4o:Hierarchy}, and related to dimensions with properties \texttt{qb4o:inDimension} and   \texttt{qb4o:hasHierarchy}. Levels are connected with the hierarchies they belong  using the property \texttt{qb4o:hasLevel}.
 
  \item Hierarchies are composed of pairs of levels, which are represented using  the class \texttt{qb4o:HierarchyStep}. Pairs are related to hierarchies via the property \\
   \texttt{qb4o:inHierarchy}. For each pair we distinguish the role of each level using the properties  \texttt{qb4o:childLevel} and
\texttt{qb4o:parentLevel}.The property \texttt{qb4o:rollup} is used to relate each pair with the property that implements the RUP relation for that step, which is an instance of \texttt{qb4o:RollupProperty}. The cardinality of this relationship (1 to 1, 1 to N, etc.) is stated using the \texttt{qb4o:pcCardinality} property 
and instances of the \texttt{qb4o:Cardinality} class. 
 \end{itemize}
\vspace{-0.2cm} 
\qed
\end{definition}

\begin{example} (\textsf{Citizenship} dimension schema in QB4OLAP) 
\label{ex:dimScQB4OLAP}
We show next the representation in QB4OLAP of
the schema of the \textsf{Citizenship} dimension of Example \ref{ex:dimSc}. 
\begin{lstlisting}[style=tiny]
@prefix property: <http://eurostat.linked-statistics.org/property#> . 
@prefix schema: <http://www.fing.edu.uy/inco/cubes/schemas/migr_asyapp#> .
@prefix qb4o: <http://purl.org/qb4olap/cubes#> .

schema:citizenshipDim a qb:DimensionProperty ;
  rdfs:label "Applicant citizenship dimension"@en;
  qb4o:hasHierarchy schema:citizenshipGeoHier, schema:citizenshipGovHier.

# dimension hierarchies
schema:citizenshipGeoHier a qb4o:Hierarchy ;
   rdfs:label "Applicant citizenship Geo Hierarchy"@en ;
   qb4o:inDimension schema:citizenshipDim;
   qb4o:hasLevel property:citizen, schema:continent.

schema:citizenshipGovHier a qb4o:Hierarchy ;
   rdfs:label "Applicant citizenship Government Hierarchy"@en ;
   qb4o:inDimension schema:citizenshipDim;
   qb4o:hasLevel property:citizen, schema:governmentType.

#hierarchy levels and attributes
property:citizen a qb4o:LevelProperty;
  rdfs:label "Country of citizenship"@en;
  qb4o:hasAttribute schema:countryName.
schema:countryName  a qb4o:LevelAttribute;
    rdfs:label "Country name"@en ; rdfs:range xsd:string.
     
schema:continent a qb4o:LevelProperty;
  rdfs:label "Continent"@en;
  qb4o:hasAttribute schema:continentName.
schema:continentName  a qb4o:LevelAttribute;
    rdfs:label "Continent name"@en ; rdfs:range xsd:string.

schema:governmentType a qb4o:LevelProperty;
  rdfs:label "Government Type"@en;
  qb4o:hasAttribute schema:governmentName .
schema:governmentName  a qb4o:LevelAttribute;
    rdfs:label "Government type name"@en ;rdfs:range xsd:string.

#rollup relationships
schema:inContinent a qb4o:RollupProperty.
schema:hasGovType a qb4o:RollupProperty.  

#hierarchy steps
_:ih43 a qb4o:HierarchyStep;
  qb4o:inHierarchy schema:citizenshipGeoHier;
  qb4o:childLevel property:citizen;
  qb4o:parentLevel schema:continent;
  qb4o:pcCardinality qb4o:OneToMany;
  qb4o:rollup schema:inContinent.

_:ih44 a qb4o:HierarchyStep;
  qb4o:inHierarchy schema:citizenshipGovHier;
  qb4o:childLevel property:citizen;
  qb4o:parentLevel schema:governmentType;
  qb4o:pcCardinality qb4o:OneToMany;
  qb4o:rollup schema:hasGovType.
\end{lstlisting}
\end{example}

\begin{definition}(\textbf{Dimension instance in QB4OLAP})
A dimension instance in QB4OLAP is an RDF graph that uses the terms defined in QB 
and QB4OLAP vocabularies, and also the IRIs defined to represent the dimension schema, as follows:
 (a) Each level member is represented by an IRI and   is related to each level it belongs to,
  via the \texttt{qb4o:memberOf} property. (b) For each step, the $\mathsf{RUP}$ relation between level members is
represented using the property linked to the step in the schema via \texttt{qb4o:rollup} property.
\vspace{-0.2cm} 
\qed
\label{def:dimInsQB4O}
\end{definition}

\begin{example} (\textsf{Citizenship} dimension instance in QB4OLAP) 
\label{ex:dimInsQB4OLAP}
Below we show part of the triples in the  QB4OLAP representation 
of the instance of the \texttt{Citizenship} dimension in Example \ref{ex:dimIns}.  
Note that some triples are enriched with links to external data, like DBpedia.  
 \begin{lstlisting}[style=tiny]
@prefix citizen: <http://eurostat.linked-statistics.org/dic/citizen#> .
@prefix citDim: <http://www.fing.edu.uy/inco/cubes/dims/migr_asyapp/citizen#> .
@prefix dbpedia:	<http://dbpedia.org/resource/> .

citizen:AD
  qb4o:memberOf property:citizen ;
  schema:inContinent citizen:EU ;
  schema:hasGovType dbpedia:Unitary_state ;
  skos:prefLabel "Andorra"@de , "Andorra"@en , "Andorre"@fr .       
citizen:ZW
  qb4o:inLevel property:citizen ;
  schema:inContinent citizen:AF ;
  schema:hasGovType dbpedia:Presidential_system;
  skos:prefLabel "Simbabwe"@de , "Zimbabwe"@fr , "Zimbabwe"@en .
citizen:EU
  qb4o:inLevel schema:continent ;
  skos:prefLabel "Europe"@en .  
citizen:AF
  qb4o:inLevel schema:continent ;
  skos:prefLabel "Africa"@en . 
dbpedia:Unitary_state
  qb4o:inLevel schema:governmentType ;
  skos:prefLabel "Unitary state"@en .
...
\end{lstlisting}
\end{example}

\begin{definition}(\textbf{Cube schema in QB4OLAP})
A cube schema in QB4OLAP is represented as an instance of 
the \texttt{qb:DataStructureDefinition} class. The dimensions and measures in the cube are  represented  using the  properties \texttt{qb:measure} and \texttt{qb:dimension}. The cardinality between
observations and dimensions is represented using the property \texttt{qb4o:cardinality}.
Aggregate functions are represented via the \texttt{qb4o:AggregateFunction}, and a property is defined to associate measures with aggregate functions
(\texttt{qb4o:aggregateFunction}). This property,
 allows a given measure to be associated with different 
aggregate functions in different cubes. 

\qed
\label{def:cubeSchQB4O}
\end{definition}

\begin{example} (Cube schema in QB4OLAP) 
\label{ex:cubeSchQB4O}
We show below   the QB4OLAP representation 
of the \textsf{Asylum\_application} cube schema.  
\begin{lstlisting}[style=tiny]
schema:migr_asyappctzmCUBE
      rdf:type qb:DataStructureDefinition ;
      qb:component [ qb:measure sdmx-measure:obsValue; qb4o:aggregateFunction qb4o:sum ] ;
      qb:component [ qb:dimension schema:sexDim ; qb4o:cardinality qb4o:ManyToOne] ;
      qb:component [ qb:dimension schema:ageDim ; qb4o:cardinality qb4o:ManyToOne];
      qb:component [ qb:dimension schema:timeDim ; qb4o:cardinality qb4o:ManyToOne];
      qb:component [ qb:dimension schema:asylappDim ; qb4o:cardinality qb4o:ManyToOne] ;
      qb:component [ qb:dimension schema:citizenshipDim ; qb4o:cardinality qb4o:ManyToOne ] ;
      qb:component [ qb:dimension schema:destinationDim ; qb4o:cardinality qb4o:ManyToOne] ;
      skos:notation "migr_asyappctzmCUBE" . 
\end{lstlisting}
\end{example}

\begin{definition}(\textbf{Cuboid instance in QB4OLAP})
A cuboid instance in QB4OLAP is a set of \texttt{qb:Observations} organized in a 
\texttt{qb:DataSet}. The set of levels of the cuboid is  an instance of the \texttt{qb:DataStructureDefinition} class, and 
the property \texttt{qb:structure} is used to relate them. To state the fact 
 that a  cuboid instance adheres to a specific cube schema we use
  the property \texttt{qb4o:isCuboidOf}.
\qed
\label{def:cuboidInstanceQB4O}
\end{definition}

\begin{example} (Cuboid instance in QB4OLAP) 
\label{ex:cuboidInstanceQB4O}
Let us consider the cuboid that corresponds to the schema of 
Example~\ref{ex:cubeSchQB4O}, which considers the lowest level for each
dimension in the cube (i.e., the bottom of the lattice). 
Below  we show  the QB4OLAP representation 
of the set of levels of the cuboid, the definition of a dataset that represents 
the cuboid instance, and also a
cell in this cuboid (\texttt{qb:Observation}), which corresponds to the last row in Table  \ref{fig:cuboidInstance}.

 \begin{lstlisting}[style=tiny]
@prefix eurostat: <http://eurostat.linked-statistics.org/data/>
@prefix cell: <<http://eurostat.linked-statistics.org/data/migr_asyappctzm#>
 
schema:migr_asyappctzmBOTTOM
  rdf:type qb:DataStructureDefinition ;
  qb4o: isCuboidOf schema:migr_asyappctzmCUBE;
  qb:component [ qb:measure sdmx-measure:obsValue; qb4o:aggregateFunction qb4o:sum ] ;
  qb:component [ qb4o:level property:age ] ;
  qb:component [ qb4o:level sdmx-dimension:refPeriod ] ;
  qb:component [ qb4o:level property:sex ] ;
  qb:component [ qb4o:level property:geo ] ;
  qb:component [ qb4o:level property:citizen] ;
  qb:component [ qb4o:level property:asyl_app ] ;
  skos:notation "migr_asyappctzmBOTTOM" .

eurostat:migr_asyappctzm 
  rdf:type qb:DataSet;
  qb:structure schema:migr_asyappctzmBOTTOM. 
  
cell:M,CD,F,Y18-34,NASY_APP,BE,2013M03
  rdf:type qb:Observation;
  qb:dataSet eurostat: migr_asyappctzm;
  property:citizen citizen:CD;
  property:sex sex:F;
  property:age age:Y18-34;
  property:asyl_app asyl_app:NASY_APP;
  property:geo geo:BE;
  measure:obsValue 30;
  sdmx-dimension:refPeriod time:201303.  
\end{lstlisting}
\end{example}

\out{
According to Definition \ref{def:cubeIns}, a cube instance is the lattice 
$\lbrace \mathsf{CB},\preceq \rbrace$ where \textsf{CB} is the set of all   
possible cuboids that adhere to a cube schema, and $\preceq$ is the order 
between adjacent cuboids in \textsf{CB}. For each pair of adjacent 
cuboids $\mathsf{Cb}_1 \preceq \mathsf{Cb}_2$, each cell
in $ \mathsf{Cb}_2$ can be computed from the cells in 
$\mathsf{Cb}_1$. Therefore, starting from the bottom cuboid  in the lattice 
(the cuboid instance whose cells are  members of the bottom levels
in each dimension of the  schema), all the possible cuboids that form the
cube instance can be computed  iteratively. In a QB4OLAP representation,   adjacent 
cuboids can be computed using SPARQL queries. Algorithm \ref{alg:cubeInstance} 
performs this computation, as explained below.

Algorithm \ref{alg:cubeInstance} takes as input a cuboid instance $\mathsf{Cb}_1$ 
represented in QB4OLAP and a level $l_p$ such that $l_p \notin \mathcal{V}_{Cb_1}$
and $\exists l_c \in \mathcal{V}_{Cb_1}$, where $l_c \rightarrow l_p$ holds in 
the lattice $(\mathcal{L}, \rightarrow)$ 
of a dimension $D_k$; the algorithm  produces a SPARQL query 
that computes a cuboid instance $\mathsf{Cb}_2$ 
that satisfies $\mathsf{Cb}_1 \preceq \mathsf{Cb}_2$ (i.e. $\mathcal{V}_{Cb_1} 
-\mathcal{V}_{Cb_2} = \lbrace l_c \rbrace$ and $\mathcal{V}_{Cb_2} 
-\mathcal{V}_{Cb_1} = \lbrace l_p \rbrace$).
To improve the clarity of the presentation, we use the auxiliary functions 
defined in Table\ref{tab:aux2} and also assume that it 
is possible to access and modify different parts of a SPARQL query via the properties: \textit{resultFormat}, 
\textit{grPattern}, and \textit{groupBy}, among others and that $add(s)$ 
appends $s$ to a particular part of the query. We will use this algorithm in the next section.
}

\out{
\begin{algorithm}
\scriptsize
\caption{Generates a SPARQL query that computes $\mathsf{Cb}_2$, such that 
$\mathsf{Cb}_1 \preceq \mathsf{Cb}_2$}
\label{alg:cubeInstance}
  \begin{algorithmic}[1]
\Require{$\mathsf{Cb}_1$ is a cuboid instance in QB4OLAP, $l_p$ is a level such 
that $l_c \in \mathcal{V}_{Cb_1}$ and $l_c \rightarrow l_p$ in the lattice 
$(\mathcal{L}, \rightarrow)$ 
of a dimension $D_k$ }
\Ensure $query$ is a SPARQL CONSTRUCT query that represents a cuboid instance 
$\mathsf{Cb}_2$ 
that satisfies $\mathsf{Cb}_1 \preceq \mathsf{Cb}_2$
    \Statex
\Function{CreateAdjacentCuboidInstance}{}
  \Let{$l_c$}{getLevel($D_k$)}
  \ForAll{$l\in L= levels(\mathsf{Cb}_1)$}
    \If{$l \neq l_c$}
    \State newVar($l_i$)
    \State query.resultFormat.add(?id , l, val($l_i$))
    \State sq.resultFormat.add(val($l_i$))
    \State sq.grPattern.add(?i , l, val($l_i$))
    \State sq.groupBy.add(val($l_i$))
    \EndIf
  \EndFor 
\ForAll{$m\in M = measures(\mathsf{Cb}_1)$} 
  \State f = aggFunction(m)
  \State newVar($m_i$); newVar($ag_i$)
  \State query.resultFormat.add(?id , m, val($ag_i$))
  \State sq.resultFormat.add(f(val($m_i$)) AS $ag_i$)
  \State sq.grPattern.add(?i , m, val($m_i$))
\EndFor
\State newVar($lm_i$),newVar($plm_i$)
\State sq.grPattern.add(?i , val($l_c$), val($lm_i$))
\State sq.grPattern.add(val($plm_i$), \texttt{qb4o:inLevel}, val($l_p$))
\State sq.grPattern.add(val($lm_i$), \texttt{skos:broader},val($plm_i$))
\State sq.groupBy.add($plm_i$)
\State sq.resulFormat.add($plm_i$)
\State query.resultFormat.add(?id , $l_p$, val($plm_i$))
\State query.grPattern.set(sq)
\State \Return{$query$} 
\EndFunction
  \end{algorithmic}
\end{algorithm}
}

\section{Querying Data Cubes}
\label{sec:queries}

We now present our proposal for querying data cubes on the semantic web.
 We follow  the approach presented in 
\cite{CCG+13}, 
where a clear separation between the conceptual and the logical 
levels is made.   In this way, we formally define a collection of 
operators, whose semantics is clearly defined using the data model of 
Section \ref{sec:cubes:formal}. This set of operators conforms our 
query language (QL) at the conceptual level. The user then will write her 
queries at this level, and they will be translated into a 
SPARQL query over the QB4OLAP-based RDF representation (at the logical level) using the algorithms presented in Section \ref{sec:queries:SPARQL} .

\subsection{Data Cube Algebra}
\label{sec:queries:algebra}

The operators proposed in the Cube Algebra sketched in~\cite{CCG+13}  can be 
classified into two groups: (1) Operations that navigate  
the cube instance to which the input cuboid belongs, which are called 
\textit{instance preserving operations (IPO)}; 
and (2) Operations that generate a new cube instance, which are called 
\textit{instance generating operations (IGO)}. The \textsc{Roll-up} and 
\textsc{Drill-down} operations belong to the first group, while \textsc{Slice}, 
\textsc{Dice} and \textsc{Drill-across} belong
to the second. We now define each of the operations and provide their semantics.

\subsubsection{Instance Preserving Operations (IPO)}

IPOs take as input a cuboid in a cube instance, and 
return another cuboid in the \textit{same} instance. We consider
the following sets: \textbf{C} is the set of all the cuboids in a cube 
instance, 
\textbf{D} is the set of dimensions, \textbf{M} is the set of measures, 
\textbf{L} is the set of dimension levels, and \textbf{B} is the set of boolean 
expressions over level attributes and measures. For clarity, and to simplify the definitions, 
we assume that the aggregate function associated to the measures is \textsc{SUM}, so we 
drop $\mathcal{F}$ from the cube schema definition. 

The \textbf{\textsc{Roll-up}} operator is a function \textsc{Roll-up}: $C 
\times 
D \times L \rightarrow C$ that summarizes data at a higher level in a dimension 
hierarchy.
It is defined as follows:

\begin{definition} (\textbf{ROLL-UP Operator}). Given a cube instance 
\textsf{CI}
with schema $\langle \mathsf{CS}, \mathcal{D}_{in}, \mathcal{M}_{in} \rangle$,
a cuboid $\mathsf{C}_{in} \in \mathsf{CI}$ with its correspondent set of levels 
$\mathcal{V}_{C_{in}}$, 
a dimension $ \mathsf{D} \in \mathcal{D}_{in}$ with schema $\langle \mathsf{D}, 
\mathcal{L}, \rightarrow \rangle$, 
and two levels $\mathsf{l}_{in}, \mathsf{l}_{out}$ in $\mathcal{D}_{in}$ such 
that: $\mathsf{l}_{in} \in \mathcal{V}_{C_{in}}$ 
and $\mathsf{l}_{in} \rightarrow^* \mathsf{l}_{out}$ in $\langle \mathcal{L}, 
\rightarrow \rangle$, then \textsc{Roll-up}$(\mathsf{C}_{in},\mathcal{D}_{in}, 
\mathsf{l}_{out})$
returns a cuboid $\mathsf{C}_{out} \in \mathsf{CI}$ such that 
$\mathcal{V}_{C_{out}}=(\mathcal{V}_{C_{in}}-\lbrace \mathsf{l}_{in} \rbrace) 
\cup \lbrace \mathsf{l}_{out} \rbrace$. 
Notice that $\mathsf{C}_{in} \prec \mathsf{C}_{out}$ in the lattice 
\textsf{CI}. 
\qed 
\label{def:rollup}
\end{definition}

The \textbf{\textsc{Drill-down}} operator is a function \textsc{Drill-down}: $C 
\times D \times L \rightarrow C$ that disaggregates data down to a specific 
level in a dimension hierarchy. 

\begin{definition} (\textbf{DRILL-DOWN Operator}). Given a cube instance 
\textsf{CI}
with schema $\langle \mathsf{CS}, \mathcal{D}_{in}, \mathcal{M}_{in} \rangle$,
a cuboid $\mathsf{C}_{in} \in \mathsf{CI}$ with its correspondent set of levels 
$\mathcal{V}_{C_{in}}$, 
a dimension $ \mathsf{D} \in \mathcal{D}_{in}$ with schema $\langle \mathsf{D}, 
\mathcal{L}, \rightarrow \rangle$, 
and two levels $\mathsf{l}_{in}, \mathsf{l}_{out}$ in $\mathcal{D}_{in}$ such 
that: $\mathsf{l}_{in} \in \mathcal{V}_{C_{in}}$ 
and $\mathsf{l}_{out} \rightarrow^* \mathsf{l}_{in}$ in $\langle \mathcal{L}, 
\rightarrow \rangle$, then 
\textsc{Drill-down}$(\mathsf{C}_{in},\mathcal{D}_{in}, \mathsf{l}_{out})$
returns a cuboid $\mathsf{C}_{out} \in \mathsf{CI}$ such that 
$\mathcal{V}_{C_{out}}=(\mathcal{V}_{C_{in}}-\lbrace \mathsf{l}_{in} \rbrace) 
\cup \lbrace \mathsf{l}_{out} \rbrace$. 
Notice that $\mathsf{C}_{out} \prec \mathsf{C}_{in}$ in the lattice 
\textsf{CI}. 
\qed 
\label{def:drilldown}
\end{definition}

It is straightforward to show, using the lattice of cuboids, 
that the cuboid produced by a \textsc{Drill-down} on a dimension D is always reachable from
the bottom of the lattice, so it can also be obtained performing a 
\textsc{Roll-up} over the same dimension D from the bottom cuboid. 
We will use this result in the sequel.

\subsubsection{Instance Generating Operations (IGO)}

IGOs generate a new cube instance, which
may have the same schema than the original one (e.g. in the case of \textsc{Dice}), or may 
have a different schema (e.g. \textsc{Slice} and \textsc{Drill-across}). In all the 
cases, these operations take as input a cuboid in a cube instance, and return a 
cuboid in another instance or cuboid lattice, induced by the application of the 
operation.

The \textbf{\textsc{Dice}} operator is a function \textsc{Dice}: $C \times B 
\rightarrow C$ that selects the values in dimension levels and measures that 
satisfy a boolean condition. It resembles
the \textsc{Selection} ($\sigma$) operation in relational algebra.

\begin{definition} (\textbf{DICE Operator}). Given a cube instance \textsf{CI}
with schema $\langle \mathsf{CS}, \mathcal{D}_{in}, \mathcal{M}_{in} \rangle$,
a cuboid $\mathsf{C}_{in} \in \mathsf{CI}$ with its correspondent set of levels 
$\mathcal{V}_{C_{in}}$, and a boolean condition $\phi$ over the measures in 
$\mathcal{M}_{in}$ and/or the attributes of the levels in 
$\mathcal{V}_{C_{in}}$, \textsc{Dice}$(\mathsf{C}_{in},\phi)$ returns a cuboid 
$\mathsf{C}_{out} \in C$ as follows: \\
(a) $c_i = (c_{i_1}, \dots, c_{i_n}, m_{i_1},\dots m_{i_s}) \in 
\mathsf{C}_{out}$ \newline 
if $\exists c_j = (c_{j_1}, \dots, c_{j_n}, m_{j_1},\dots m_{j_s}) \in 
\mathsf{C}_{in}$ and 
$c_{i_p}=c_{j_p}$ $\forall p, p=1,\dots,n$, $m_{i_q}=m_{j_q}$ $\forall q, 
q=1,\dots,s$, and 
$c_j$ satisfies $\phi$; \\
(b) $\mathcal{V}_{C_{out}} = \mathcal{V}_{C_{in}}$
\qed
\label{def:dice}
\end{definition}

The \textbf{\textsc{Slice}} operator is a function \textsc{Slice}: $C \times (D 
\cup M) \rightarrow C$ that reduces the dimensionality of a cube by removing 
one of its dimensions or measures.
In the case of eliminating a 
dimension, the \textsc{Roll-up} operation is applied to this dimension in the 
cuboid before removing it.

\begin{definition} (\textbf{SLICE Operator}). Given a cube instance \textsf{CI}
with schema $\langle \mathsf{CS}, \mathcal{D}_{in}, \mathcal{M}_{in} \rangle$, 
where $\mid \mathcal{D}_{in}\mid > 1$ or where $\mid \mathcal{M}_{in}\mid > 1$ 
a 
cuboid $\mathsf{C}_{in} \in \mathsf{CI}$ with its correspondent set of levels 
$\mathcal{V}_{C_{in}}$, 
and a dimension $D \in \mathcal{D}_{in}$ or a measure $M \in \mathcal{M}_{in}$, 
according to the input parameters: \\
(1) \textsc{Slice}$(\mathsf{C}_{in},D)$ returns a cuboid $\mathsf{C}_{out} \in 
C$ as follows: \\
(a) $c_i = (c_{i_1}, \dots, c_{i_{k-1}},c_{i_{k+1}}, m_{i_1},\dots m_{i_s}) 
\in \mathsf{C}_{out}$ \newline
if $\exists c_j = (c_{j_1}, \dots,c_{i_{k-1}},\text{all},c_{i_{k+1}} \dots, 
c_{j_n}, m_{j_1},\dots m_{j_s}) \in $\textsc{Roll-up}$(\mathsf{C}_{in},D,All)$ and 
$c_{i_p}=c_{j_p}$ $\forall p, p=1,\dots,n$, $p \neq k$, and
$m_{i_q}=m_{j_q}$ $\forall q, q=1,\dots,s$; \\
(b) $\mathcal{V}_{C_{out}} = \mathcal{V}_{C_{in}}- \lbrace l_d \rbrace$, 
where 
$l_d$ is the level corresponding to dimension $D$ in $\mathsf{C}_{in}$. \\
(2) \textsc{Slice}$(\mathsf{C}_{in},M)$ returns a cuboid $\mathsf{C}_{out} \in 
C$ as follows: \\
(a) $c_i = (c_{i_1}, \dots, c_{i_n}, 
m_{i_1},\dots,m_{i_{k-1}},m_{i_{k+1}},\dots,m_{i_s}) \\
 \in \mathsf{C}_{out}$ 
~if $\exists c_j = (c_{j_1},\dots, c_{j_n}, 
m_{j_1},\dots,m_{j_{k-1}},m_{j_{k}},\\
m_{j_{k+1}}, \dots, m_{j_s}) \in 
\mathsf{C}_{in}$ and 
$c_{i_p}=c_{j_p}$ $\forall p, p=1,\dots,n$, and
$m_{i_q}=m_{j_q}$ $\forall q, q=1,\dots,s$, $q \neq k$; \\
(b) $\mathcal{V}_{C_{out}} = \mathcal{V}_{C_{in}}$
\qed
\label{def:slice}
\end{definition}

The \textbf{\textsc{Drill-across}} operator is a function 
\textsc{Drill-across}: 
$C \times C \rightarrow C$ that performs the union of two cuboids
that are defined over the same dimensions, and contain the same instance, but 
differ in the measures. It allows to compare measures from different cuboids 
and resembles the \textsc{Join} ($\Join$) operation in relational algebra.
In this paper we will not address this operator, and limit ourselves to 
unary operations, that means, operations over single data cubes. Thus, we omit 
the formal definition of the operation.

\subsection{Algebra Operations as SPARQL Queries}
\label{sec:queries:SPARQL}

\begin{table}[t]
\caption{Auxiliary functions}
\label{tab:aux2}
\scriptsize
\begin{tabularx}{.9\linewidth}{|l|X|}

\hline
\textbf{Function signature} & \textbf{Description} \\\hline
newVarName() & Generates and returns a unique SPARQL variable name.\\\hline

val(v) & Returns the value stored in variable v\\\hline

levels(s) & Returns all the levels in a schema s (i.e., all the values of ?l 
that satisfy  s \texttt{qb:component} ?c. ?c \texttt{qb4o:level} ?l)\\\hline

getLevel(s,d) & Returns the only level l that corresponds to dimension d in the schema s (i.e. the only value of ?l 
that satisfies  s \texttt{qb:component} ?c. ?c \texttt{qb4o:level} ?l. ?h \texttt{qb4o:hasLevel} ?l. ?h \texttt{qb4o:inDimension} ?d)\\\hline

levelsPath($l_o$,$l_d$)& Returns a levels path from level $l_o$ to $l_d$ \\\hline
getRollup($l_c$,$l_p$)& Resturns the predicate that implements the RUP function from level $l_c$ to level $l_p$ \\\hline
measures(s) & Returns all the measures in a schema s (all the values of ?m that satisfy s 
\texttt{qb:component} ?c. ?c \texttt{qb:measure} ?m)\\\hline
aggFunction(m,s)& Returns the aggregation function of measure m (all the values 
of ?f that satisfy s \texttt{qb:component} ?c. ?c \texttt{qb:measure} ?m 
;\texttt{qb4o:aggregateFunction} ?f).\\\hline
\end{tabularx}
\end{table}

In this section we show  how the operators above can be implemented as 
SPARQL queries over QB4OLAP-based RDF data cubes. 
Before that, we would like to discuss on the result form of these queries.

SPARQL queries may return results in different formats. In particular \texttt{SELECT} queries return a table of values, while \texttt{CONSTRUCT} queries return a graph (i.e., a set of triples). Since each operator returns a cuboid in a certain cube instance, and according to Definitions \ref{def:cubeSchQB4O}
and \ref{def:cuboidInstanceQB4O}, cuboids in QB4OLAP are RDF graphs, it is evident that algebra operators should be implemented in SPARQL using \texttt{CONSTRUCT} queries. Despite this, SPARQL 1.1 does not allow to compute aggregations in a \texttt{CONSTRUCT} query, and therefore our approach is to use subqueries. We then produce two queries: (a) an inner \texttt{SELECT} query to compute aggregations, and (b) an outer \texttt{CONSTRUCT} query that generates the graph using the computed results. Notice that the inner query is responsible for the actual computation of values, while the outer query just generates the output as a graph.

We now present the algorithms that generate the SPARQL implementation of each operator.
To improve the clarity of the presentation, we use the auxiliary functions 
defined in Table \ref{tab:aux2}. We also use an abstract representation of a SPARQL query, where for each query:
(a) \textit{queryType} can be \texttt{SELECT} or \texttt{CONSTRUCT},
(b) \textit{resultFormat}  represents the set of variables and expressions included in the \texttt{SELECT} clause or the set of BGPs included in the \texttt{CONSTRUCT} clause, depending on the type of the query,
(c) \textit{grPatterns} represents the set of graph patterns in the \texttt{WHERE} clause, 
(d) \textit{subQueries} represents the set of subqueries in the \texttt{WHERE} clause, 
(e) \textit{filter} represents a \texttt{FILTER} clause,
(f) and \textit{groupBy} represents the set of variables included in the \texttt{GROUP BY} clause. 
We assume that each of these parts
can be accessed and modified, and we use the dot notation (``.") to access them. We also consider a function $add()$, such that $add(s)$ appends $s$ to a particular part of the query. For example, given a query $q$  such that $q.queryType == $``SELECT'' the instruction $q.resultFormat.add(v)$ adds the variable $v$ to the \texttt{SELECT} clause of $q$.

\subsubsection{IPO Operations as SPARQL Queries}
\label{sec:queries:SPARQL:IPO}

According to Definition~\ref{def:cubeIns}, a cube instance is the lattice 
$\lbrace \mathsf{CB},\preceq \rbrace$ where \textsf{CB} is the set of all   
possible cuboids that adhere to a cube schema, and $\preceq$ is the order 
between adjacent cuboids in \textsf{CB}. As stated by Definition~\ref{def:cuboidAdj} 
for each pair of adjacent 
cuboids $\mathsf{Cb}_1 \preceq \mathsf{Cb}_2$, each cell
in $ \mathsf{Cb}_2$ can be computed from the cells in 
$\mathsf{Cb}_1$. Therefore, starting from the bottom cuboid  in the lattice, which is 
the cuboid instance whose cells are  members of the bottom levels
in each dimension of the  schema, all the possible cuboids that form the
cube instance can be computed  incrementally. 

To compute the \textsc{Roll-up} operation over a cuboid $\mathsf{Cb_{in}}$ and 
a dimension $D$  it 
suffices to start at $\mathsf{Cb_{in}}$, and navigate the cube 
lattice visiting adjacent cubes that differ only in 
the level associated to dimension $D$, until we reach a cuboid $\mathsf{Cb_{out}}$ that has the 
desired level in dimension $D$ (this path is unique).

\begin{remark} It is not necessary to compute all the cuboids in the path, it suffices to compute the target cuboid. To do so it is necessary to add all the triples needed to traverse the dimension hierarchy up to the target level and aggregate measure values up to this level.
\qed
\end{remark}

As already mentioned, two SPARQL queries are needed: an inner query $q_{in}$ that traverses the dimension hierarchy and computes aggregate values using \texttt{GROUP BY}, 
and an outer one, called 
$q_{out}$ that builds triples based on the values computed in $q_{in}$.  

Algorithm \ref{alg:rollup} builds both queries simultaneously, using the $add$ function. 
Lines 2 and 3 state the query type for each query.
Line 4 states that generated observations belong to the dataset $newDS$.
Lines 6 through 12 project the members of each level in the schema into the 
result of both queries, also adding triples to the \texttt{WHERE} clause of the inner 
query  and adding the variables that represent the level members  to the \texttt{GROUP 
BY} clause also in the inner query. Lines 13 through 19 do the same for measures.  In 
Lines 14 and 18,  $f$ represents the SPARQL function corresponding to the 
aggregate  function for each measure,  and $f(val(m_i))$ is the string that 
should be included to calculate the aggregated value (e.g 
$\mbox{\textsc{sum}}(?m)$ if $val(m_i)= ?m$). Lines 20 to 33 add the triples 
needed to navigate the dimension hierarchy.  Line 22 retrieves the RDF property that implements the RUP relation for each step in the path. Line 24 adds  to the inner query, a 
triple  that associates the level member with the observation (only for the base level $l_c$ in dimension D);  Line 27 adds a triple that allows us to state to which level the level member belongs, and line 
29 retrieves the parent level member of the current level applying the RUP function obtained in Line 22 (this is done for all the levels in the path except for the target level $l_{out}$. When level $l_{out}$ is reached Lines 31 and 32 add the target level to the \texttt{GROUP BY} and \texttt{SELECT} clauses of the inner query, respectively, while Line 33 adds the target level to the outer query result.
Finally, Line 34 sets the inner query as a subquery within the \texttt{WHERE} clause of the outer query, which is returned in Line 35.  
For clarity, we have omitted the clause that generates the expression that binds 
variable $?newObs$ to a dynamically generated IRI from the values in the observation.

\begin{algorithm}[t]
\scriptsize
\caption{Generates a SPARQL query that implements a \textsc{Roll-up} in QB4OLAP}
\label{alg:rollup}
\begin{algorithmic}[1]
\Require{$\mathsf{C}_{in}$ is a cuboid instance in QB4OLAP, where $\mathcal{V}_{C_{in}}$ is the set of levels of the cuboid and $d_r$ is the dataset that represents the cuboid instance, $l_{out}$ is a level such 
that $l_c \in \mathcal{V}_{C_{in}}$ and $l_c \rightarrow^* l_{out}$ in the lattice $(\mathcal{L}, \rightarrow)$ 
of a dimension $\mathcal{D}$ }
\Ensure $q_{out}$ is a SPARQL CONSTRUCT query that represents a cuboid instance 
$\mathsf{C}_{out}$ = \textsc{Roll-up}$(\mathsf{C}_{in},\mathcal{D}, \mathsf{l}_{out})$
    \Statex
\Function{CreateRollUpQuery}{$\mathsf{C}_{in}$, $\mathcal{D}$, $l_{out}$}
\State $q_{out}$.queryType = 'CONSTRUCT'
\State $q_{in}$.queryType = 'SELECT'
\State $q_{out}$.grPatterns.add(?newObs , \texttt{qb:dataSet} , $newDS$)
  \Let{$l_c$}{getLevel($C_{in},D$)}
  \ForAll{$l\in L= levels(C_{in})$}
    \If{$l \neq l_c$}
    \Let {$l_i$}{newVar()}
    \State $q_{in}$.grPatterns.add(?obs , l, val($l_i$))
    \State $q_{in}$.groupBy.add(val($l_i$))
    \State $q_{in}$.resultFormat.add(val($l_i$))
    \State $q_{out}$.resultFormat.add(?newObs , l, val($l_i$))
    \EndIf
  \EndFor 
\ForAll{$m\in M = measures(\mathsf{C}_{in})$} 
  \Let {f}{aggFunction(m)}
  \Let {$m_i$}{newVar()}
  \Let {$ag_i$}{newVar()}
  \State $q_{in}$.grPatterns.add(?obs , m, val($m_i$))
  \State $q_{in}$.resultFormat.add(f(val($m_i$)) AS $ag_i$)
  \State $q_{out}$.resultFormat.add(?newObs , m, val($ag_i$))
\EndFor
\ForAll {$(l_i,l_j)\in path = levelsPath(l_c,l_{out})$} 
  \Let {$lm_i$}{newVar()}
  \Let {rup}{getRollup($l_i,l_j$)}
  \If{$l_i = l_c$}
    \State $q_{in}$.grPatterns.add(?obs , val($l_i$), val($lm_i$))
  \Else
   \Let {$plm_i$}{newVar()}
    \State $q_{in}$.grPatterns.add(val($plm_i$), \texttt{qb4o:memberOf}, val($l_i$))
    \If{$l_i \neq l_{out}$}
      \State $q_{in}$.grPatterns.add(val($lm_i$),rup,val($plm_i$))
    \Else
     	\State $q_{in}$.groupBy.add($plm_i$)
		\State $q_{in}$.resulFormat.add($plm_i$)
		\State $q_{out}$.resultFormat.add(?newObs , $l_p$, val($plm_i$))
    \EndIf
  \EndIf
\EndFor
\State $q_{out}$.subqueries.add($q_{in}$)
\State \Return{$q_{out}$} 
\EndFunction
  \end{algorithmic}
\end{algorithm}

\begin{example}  
The SPARQL query generated by
 Algorithm \ref{alg:rollup} for \textsc{RollUp}(Asylum\_application, Citizenship, Continent) is:  

\label{ex:rollup}
\begin{lstlisting}[style=tiny]
PREFIX property:  <http://eurostat.linked-statistics.org/property#>
PREFIX qb:      <http://purl.org/linked-data/cube#>
PREFIX sdmx-dimension:  <http://purl.org/linked-data/sdmx/2009/dimension#>
prefix sdmx-measure:   <http://purl.org/linked-data/sdmx/2009/measure#>
prefix qb4o:           <http://purl.org/qb4olap/cubes#>
prefix schema: <http://www.fing.edu.uy/inco/cubes/schemas/migr_asyapp#>
prefix queries: <http://www.fing.edu.uy/inco/cubes/queries/migr_asyapp#>

CONSTRUCT { 
  ?newObs a qb:Observation ;  qb:dataSet queries:ejRollup; 
  sdmx-dimension:refPeriod ?time ;   property:sex ?sex ;  
  property:geo ?geo;  property:age ?age;
  property:asyl_app ?apptype;  schema:continent ?citContinent ;
  sdmx-measure:obsValue ?sumApp } 
WHERE {
  SELECT ?newObs ?time ?sex ?geo ?age ?apptype ?citContinent (SUM(xsd:integer(?m)) AS ?sumApp) 
  FROM <http://www.fing.edu.uy/inco/cubes/instances/migr_asyapp_clean>
  FROM <http://www.fing.edu.uy/inco/cubes/schemas/migr_asyappctzmQB4O13>
  WHERE{
    ?obs qb:dataSet  <http://eurostat.linked-statistics.org/data/migr_asyappctzm> ; 
    sdmx-dimension:refPeriod ?time ;
    property:sex ?sex ;    property:geo ?geo ;
    property:age ?age;    property:asyl_app ?apptype;
    sdmx-measure:obsValue ?m ;    property:citizen ?citizen .
    ?citizen qb4o:memberOf property:citizen.
    ?citizen schema:inContinent ?citContinent. ?citContinent qb4o:memberOf schema:continent.
    bind (iri(concat('http://www.fing.edu.uy/inco/cubes/instances/migr_asyapp',
    md5(concat(str(?time), str(?sex), str(?geo), str(?age), str(?apptype), 
        str(?citContinent))))) as ?newObs) 
  }
  GROUP BY ?newObs ?time ?sex ?geo ?age ?apptype ?citContinent 
}
\end{lstlisting}
\vspace{-1cm}
\qed
\end{example}

In Section \ref{sec:queries:algebra} we have  discussed that it is possible to transform a \textsc{Drill-down} operation into a \textsc{Roll-up}, therefore
there is no need to provide an specific implementation for  \textsc{Drill-down} in QB4OLAP.

\subsubsection{IGO Operations as SPARQL Queries}
\label{sec:queries:SPARQL:IGO}

IGO operations take as input a cuboid in a cube instance, induce a new cube 
instance, and return a cuboid in this newly induced lattice of cuboids.
In some cases (e.g. \textsc{Slice}) the operations 
also affect the schema of the cube before producing the cube instance.

The \textsc{Dice} operation   takes as input a cuboid in a cube instance, and a boolean 
expression $\phi$ over 
measure values and/or attribute values, and returns a cuboid in a new cube instance 
keeping only the cells from the input cuboid that satisfy $\phi$. 
The implementation of this operator in SPARQL selects the \texttt{qb:Observations} that satisfy $\phi$.
Since measures and attributes are literals, conditions over them can be implemented as \texttt{FILTER} clauses.
Also, conditions that only involve equality can be efficiently implemented via graph patterns, 
restricting the result of the query to observations that are related to a particular level member. 
Inequalities over level members are represented as \texttt{FILTER} clauses.
Algorithm \ref{alg:dice} generates the SPARQL implementation of the \textsc{Dice} operator, which is also based in an 
inner query which performs the filter, and an outer query that produces the results.

\begin{algorithm}
\scriptsize
\caption{Generates a SPARQL query that implements a \textsc{Dice} in QB4OLAP}
\label{alg:dice}
\begin{algorithmic}[1]
\Require{$\mathsf{C}_{in}$ is a cuboid instance in QB4OLAP, where with its correspondent set of levels 
$\mathcal{V}_{C_{in}}$, and a boolean condition $\phi$ over the measures in 
$\mathcal{M}_{in}$ and/or the attributes of the levels in 
$\mathcal{V}_{C_{in}}$
$\mathcal{V}_{C_{in}}$ is the set of levels of the cuboid, $\phi$ is a boolean condition over the measures in 
$\mathcal{M}_{in}$ and/or the attributes of the levels in 
$\mathcal{V}_{C_{in}}$, and $d_r$ is the dataset that represents the cuboid instance}
\Ensure $q_{out}$ is a SPARQL CONSTRUCT query that represents a cuboid instance 
$\mathsf{C}_{out}$ = \textsc{Dice}$(\mathsf{C}_{in},\phi)$
    \Statex
\Function{CreateDiceQuery}{$\mathsf{C}_{in},\phi$}
\State $q_{out}$.queryType = 'CONSTRUCT'
\State $q_{in}$.queryType = 'SELECT'
\State $l_{vars}$ = []
\State $m_{vars}$ = []
\State $bgps_{filter}$ = []
\State $q_{out}$.grPatterns.add(?newObs , \texttt{qb:dataSet} , $newDS$)
  \ForAll{$l\in L= levels(C_{in})$}
    \Let {$l_i$}{newVar()}
    \State $l_{vars}[l]=l_i$
    \State $q_{in}$.grPatterns.add(?obs , l, val($l_i$))
    \State $q_{in}$.resultFormat.add(val($l_i$))
    \State $q_{out}$.resultFormat.add(?newObs , l, val($l_i$))
  \EndFor 
\ForAll{$m\in M = measures(\mathsf{C}_{in})$} 
  \Let {$m_i$}{newVar()}
  \State $m_{vars}[m]=m_i$
  \State $q_{in}$.grPatterns.add(?obs , m, val($m_i$))
  \State $q_{in}$.resultFormat.add((val($m_i$))
  \State $q_{out}$.resultFormat.add(?newObs , m, val($ag_i$))
\EndFor
\Let{$treeCond$}{parseCondition($\phi$)}
\State{procCondition($treeCond, l_{vars}, m_{vars}$, $bgps_{filter}$,$cond_{filter}$)}
\ForAll{$bgp\in bgps_{filter}$} 
	\State $q_{in}$.grPatterns.add($bgp$)
\EndFor
	\State $q_{in}$.filter.add($cond_{filter}$)
\State $q_{out}$.subqueries.add($q_{in}$)
\State \Return{$q_{out}$} 
\EndFunction

\Require{$tree$ is a binary tree that represents a boolean condition $\phi$, where internal nodes represent
boolean operators (AND,OR,NOT) and leaves represent conditions over level attributes or measure values.$l_{vars}$ 
is the set of variables that represent level members, $m_{vars}$ is the set of variables that represent measure values}

\Ensure $q_{out}$ is a SPARQL CONSTRUCT query that represents a cuboid instance  $\mathsf{C}_{out}$ = \textsc{Dice}$(\mathsf{C}_{in},\phi)$
    \Statex
\Function {procCondition}{$tree, l_{vars}, m_{vars}, bgps, filter$}
	 \If{$tree = leaf$}
	 	\If{$tree.type ="LEVEL"$}
	 	    \Let{$v$}{findVariable($tree.element,l_{vars}$)}
	 	    \Let {$la$}{newVar()}
	 	    \State bgps.add($v$ , $tree.level$, $la$)
	 	    	\Let {$filter$}{$la$, $tree.oper$, $tree.value$)}
	 	\Else
	 		\Let{$v$}{findVariable($tree.element,m_{vars}$)}
	 		\Let {$filter$}{$v$, $tree.oper$, $tree.value$)}
	 	\EndIf
	 \Else
	 \State{procCondition($tree.left, l_{vars}, m_{vars}$, $bgps_{left}$,$filter_{left}$)}
	 \State{procCondition($tree.right, l_{vars}, m_{vars}$, $bgps_{right}$,$filter_{right}$)}
	 \State bgps.add($bgps_{left}$)
	 \State bgps.add($bgps_{right}$)
	 \State filter.add($filter_{left}$,$tree.oper$,$filter_{right}$)	 	
	 \EndIf
\EndFunction
  \end{algorithmic}
\end{algorithm}

\begin{example} (SPARQL query that implements the \textsc{Dice} operation)
The operation: \\
\textsc{Dice} (Asylum\_application, ((201303<=month<=201307) $\vee$ (\#applications >80) $\wedge$ Destination.country.countryName = Belgium)) is implemented in SPARQL as follows. 
\label{ex:dice}
\begin{lstlisting}[style=tiny]
CONSTRUCT { ?o ?p ?v}
WHERE{
  SELECT ?o ?p ?v
  FROM <http://www.fing.edu.uy/inco/cubes/instances/migr_asyapp_clean>
  FROM <http://www.fing.edu.uy/inco/cubes/schemas/migr_asyappctzmQB4O13>
  WHERE{
    ?o a qb:Observation.
    ?o qb:dataSet  <http://eurostat.linked-statistics.org/data/migr_asyappctzm>.
    ?o sdmx-dimension:refPeriod ?time. 
    ?o sdmx-measure:obsValue ?m.
    ?o <http://eurostat.linked-statistics.org/property#geo> ?lm1 .
    ?lm1 <http://www.fing.edu.uy/inco/cubes/schemas/migr_asyapp#countryName> "Belgium"@en .
    ?time schema:yearMonthNum ?timeMonthNum.
    ?o ?p ?v.
    FILTER (?timeMonthNum >= 201303 && ?timeMonthNum <= 201307&& xsd:integer(?m)>80) 
  }
}
\end{lstlisting}
\vspace{-1cm}
\qed
\end{example}

The \textsc{Slice} operation comes in two flavors. In one case, it takes as input a 
cuboid instance and a dimension. In the other, it takes a cuboid instance and a 
measure.
In both cases the implementation of this operator in QB4OLAP requires the 
creation of a new schema, where the input dimension or the measure are removed.
In the case where the operation receives a dimension 
(\textsc{Slice}$(\mathsf{C}_{in},D)$), the new cube instance $\mathsf{C}_{out}$ 
is computed as (\textsc{Roll-up}$(\mathsf{C}_{in},D,\mathsf{All})$). 
The SPARQL query generation algorithm is straightforward, and we omit it here.

\section{Related Work}
\label{sec:related}

There are two main lines of research addressing OLAP analysis of SW data, namely 
(1) extracting MD data from the SW and  loading 
them into traditional MD  data management systems for OLAP analysis; and  (2) 
performing OLAP-like analysis directly over SW data, 
e.g., over MD data represented in RDF. We next discuss them in some more 
detail. 

Relevant to the first line are the works by Nebot and 
Llavori~\cite{Nebot2012} and K\"{a}mpgen and Harth~\cite{Kampgen2011}.
The former proposes a semi-automatic method for on-demand extraction of  
semantic 
data into a MD database. In this way, data could be analyzed using traditional 
OLAP techniques. The authors present a methodology for discovering 
facts in SW data, and populating a MD model with such 
facts. They assume that data are represented as an 
OWL\footnote{\url{http://www.w3.org/TR/owl2-overview/}}  ontology.  
The proposed   methodology has four main phases: (1)  Design of the  
MD schema, where the user selects the subject of analysis that corresponds to a 
concept of the ontology, and then  selects potential dimensions. 
Then, she defines the measures, which are functions over data type 
properties; (2) Identification  and extraction of  facts from the instance 
store according to the MD schema  previously designed, producing  the base 
fact table of a DW; (3) Construction of the dimension   
hierarchies based on the instance values of the fact table and the knowledge 
available in the domain ontologies (i.e., the inferred taxonomic relationships) 
and
also considering desirable OLAP 
properties for the hierarchies; 
 (4) User specification of MD queries over the DW. 
 Once queries  are  executed,   a cube is built. Then, typical OLAP operations 
can be applied over this cube.

K\"{a}mpgen and Harth~\cite{Kampgen2011}  study the extraction of statistical 
data published using the  QB vocabulary into a MD database. 
The authors propose a mapping between the concepts in QB and a MD 
data model, and implement these mappings via SPARQL queries. 
There are four main phases in the proposed methodology: (1)  
 Extraction, where the user defines relevant data sets which are 
retrieved from the web and stored in a local triple store. Then, SPARQL queries 
are performed over this triple store to retrieve metadata on the schema, as 
well 
as   data instances;  (2) Creation of  a relational representation
of the MD data model, using the metadata retrieved in the previous step, and 
the 
population of this model   with the retrieved data;  (3) Creation of 
 a MD model to allow OLAP operations over the underlying relational 
representation. Such model is expressed using XML for Analysis 
(XMLA)\footnote{\url{http://xmlforanalysis.com}}, which allows the 
serialization 
of MD models and is implemented by several OLAP clients and servers; (4) 
Specification of  queries over the DW, using OLAP client 
applications.

The proposals described above  are based on traditional MD data management
systems, thus they  capitalize the existent knowledge in this area and can  
reuse the vast  amount of available tools. However, they require 
the existence of a local DW to store SW data. This 
restriction clashes with the autonomous and highly volatile nature of web 
data sources as changes in the sources may lead not only to updates on data 
instances but also in the structure of the DW, which would 
become  hard to update and maintain. In addition, these approaches solve only 
one part of the problem, since they do not consider the possibility of 
directly querying \textit{\`{a} la} OLAP MD data over the SW. 

The second line of research tries to overcome the drawbacks of the first one, 
exploring data models and tools that allow  publishing and performing OLAP-like 
analysis directly over SW MD data.  Terms like  \textit{self-service BI}
~\cite{DBLP:journals/jdwm/AbelloDEGMNPRTVV13}, 
\textit{Situational BI}~\cite{Loser2009}, \textit{on-demand BI}, or even 
\textit{Collaborative BI}, refer to the capability of incorporating situational 
data into the decision process  with little or no intervention of programmers 
or designers. The web, and in particular the SW, is considered as a  large 
source of data that could enrich decision processes.  
Abell\'o et al.~\cite{DBLP:journals/jdwm/AbelloDEGMNPRTVV13} present 
a  framework to support self-service BI, based on the notion of 
\textit{fusion cubes}, i.e., MD cubes that can be dynamically 
extended both in their schema and their instances, and in which data and 
metadata can be  associated with quality and provenance annotations.

To support the second approach mentioned above, the RDF Data Cube 
vocabulary~\cite{Cyganiak2014} proposes 
an RDF representation for statistical data according to the SDMX information 
model. 
Although similar to traditional MD  
data models, the SDMX semantics imposes restrictions on what can 
be represented using QB.  
Etcheverry and Vaisman~\cite{Etcheverry2012b} proposed QB4OLAP,
an extension to QB that allows to represent analytical  data according
to traditional MD models, also  presenting  a preliminary  
implementation of some OLAP operators (Roll-Up, Dice, and Slice), 
using SPARQL queries over data cubes specified using QB4OLAP.

In \cite{Ibragimov2014} the authors present a framework for Exploratory OLAP 
over Linked Open Data sources, where the MD schema of the data cube is expressed in QB4OLAP and VoID. 
Based on this MD schema the system is able to query data sources, extract and 
aggregate data, and build an OLAP cube. The MD information retrieved from 
external sources is also stored using QB4OLAP. 

For an exhaustive study of the possibilities of using SW  technologies 
in OLAP, we refer the reader to the survey by Abell\'o et 
al.~\cite{DBLP:journals/tdke/Abello2015}.

\section{Conclusion and Future work}
\label{sec:conclusion}

In this work, we  presented a formal data model for data cubes, that allows the 
definition of a conceptual 
query language to manipulate data cubes, and  showed that a data cube 
represented using this  model can be published on the  semantic web using the 
QB4OLAP vocabulary.  With a  focus  
on querying and publishing data cubes on the semantic web,  we defined a 
conceptual query language for the data model previously described.
Finally, we showed how   SPARQL queries over QB4OLAP cubes can be automatically 
produced.

In future work we will concentrate on the composition of operators and 
the experimental evaluation of this 
proposal, which, to the best of our knowledge, is the first of its kind.

\bibliographystyle{abbrv}
\bibliography{corr2015}  
%
%
\end{document}